\documentclass[twocolumn,aps,prl,showpacs,preprintnumbers,amsmath,amssymb,10pt]{revtex4-1}

\usepackage{graphicx}
\usepackage{dcolumn}
\usepackage{setspace}
\usepackage{bm}
\usepackage{multirow}

\begin{document}

\title{Extracting dynamics in the fusion of neutron-rich light nuclei}

\author{R.~T. deSouza}
\email{desouza@indiana.edu}
\author{Varinderjit Singh}
\author{S. Hudan}
\affiliation{%
Department of Chemistry and Center for Exploration of Energy and Matter, Indiana University\\
2401 Milo B. Sampson Lane, Bloomington, Indiana 47408, USA}%

\author{Z. Lin}
\affiliation{%
Department of Physics and Center for Exploration of Energy and Matter, Indiana University\\
2401 Milo B. Sampson Lane, Bloomington, Indiana 47408 USA}%
{\affiliation{%
Department of Physics, Arizona State University, \\ 450 E. Tyler Mall, Tempe, AZ 85287-1504 USA}%

\author{C.~J. Horowitz}
\affiliation{%
Department of Physics and Center for Exploration of Energy and Matter, Indiana University\\
2401 Milo B. Sampson Lane, Bloomington, Indiana 47408, USA}%

\date{\today}

\begin{abstract}

The dependence of fusion dynamics on neutron excess for light nuclei is extracted. This is 
accomplished by comparing 
the average fusion cross-section at energies just above the fusion barrier for $^{12-15}$C + $^{12}$C
with measurements of the interaction cross-section from high energy collisions. The experimental
results indicate that the fusion cross-section associated with dynamics increases with increasing 
neutron excess. Calculations with a time-dependent Hartree-Fock model fail to decribe the observed trend.

\end{abstract}


\maketitle

Nuclei are extremely interesting quantal systems. Despite a limited number of constituent particles, they manifest collective dynamics.
This collective dynamics is observed in many forms including the giant multipole resonances \cite{Bertrand76}, shape coexistence \cite{Gade16}, and the quintessential case of
fission \cite{Meitner39}. Although typically associated with the structure and reactions of mid-mass and heavy nuclei, collectivity for very
light nuclei has recently been reported \cite{Morse18}. Nuclear fission and nuclear fusion provide examples in which collective
degrees of freedom undergo substantial change as the reaction proceeds. Of particular interest is the role of collectivity for
neutron-rich nuclei as for these nuclei the dependence of the dynamics on the asymmetry between the neutron and proton densities
can be probed.
Fusion reactions provide
a powerful means to assess the response of neutron-rich
nuclei to perturbation. As fusion involves the interplay of
the repulsive Coulomb and attractive nuclear potentials,
by examining fusion for an isotopic chain one probes the
neutron density distribution and how that density distribution evolves as the two nuclei approach and overlap \cite{Singh17,Vadas18, Hudan20}.
In the following manuscript we propose a
novel perspective for investigating the role
of collective dynamics in fusion. Moreover, we present the dependence of the fusion dynamics on n/p asymmetry for the first
time including the indication that for light nuclei fusion dynamics increases with increasing neutron number.

\begin{figure}
\includegraphics[scale=0.42]{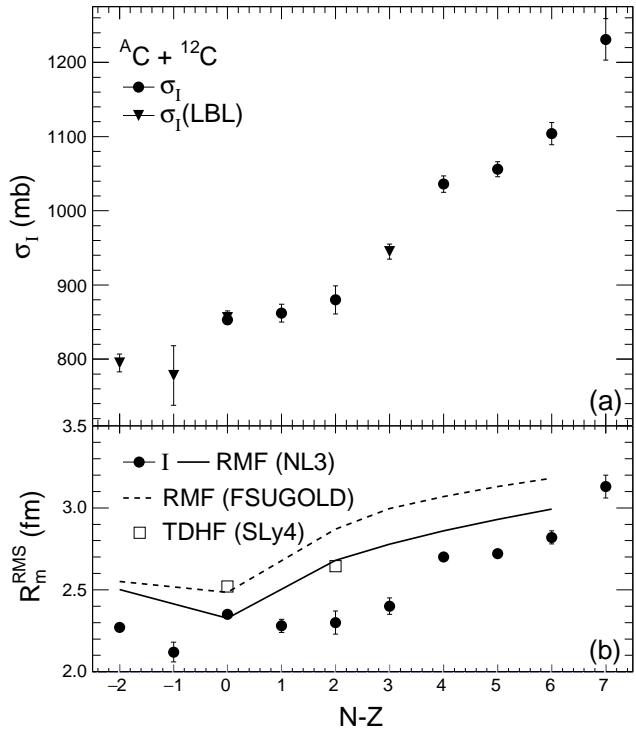}
\caption{\label{fig:fig1} Panel a): Dependence of the interaction cross-section $\sigma_{I}$ on neutron excess for various carbon isotopes. Taken from \cite{Ozawa01a}. 
Panel b): The dependence of the matter radius on neutron excess is compared with the predictions of RMF and TDHF models.
}
\end{figure}

It is well established that measurement of the interaction cross-section, $\sigma_{I}$, 
in high energy collisions is an effective means to invesitgate the
spatial extent of the matter distribution. 
At the high incident energy that these experiments were conducted at, one
expects the sudden approximation to be valid. Hence, the measured interaction 
cross-section, $\sigma_{I}$ provides a direct measure of the extent of the matter distribution. 
Systematic comparison of these cross-sections for lithium isotopes revealed the halo nature of $^{11}$Li \cite{Tanihata85a, Tanihata85b}.
Presented in Fig.~\ref{fig:fig1}a 
are the interaction cross-sections of carbon isotopes with a carbon target. Measurements for A$\geq$12
were made at E/A $\sim$ 900 MeV at GSI-Darmstadt using the high-resolution fragment separator FRS\cite{Ozawa01a, Ozawa01b} and are supplemented by the results of
earlier measurements at the LBL Bevalac\cite{Ozawa96}. The overall trend observed is an approximately linear
increase in $\sigma_{I}$ with neutron excess, (N-Z).

Closer examination of Fig.~\ref{fig:fig1}a 
provides an indication of the impact of shell structure on $\sigma_{I}$.
The dependence of $\sigma_{I}$ on neutron excess for 12$\le$A$\le$14 is weak as is the dependence for
16$\le$A$\le$18. Between $^{14}$C and $^{16}$C one observes a jump in $\sigma_{I}$ from a value of $\sim$850 mb to
$\sim$1050 mb. This increase reflects the completion of the 1p$_{\frac{1}{2}}$ with N=8 and the population of the sd-shell indicating
that the shell structure of the neutron-rich isotopes is observable through
measurement of $\sigma_{I}$ for an isotopic chain.

Through comparison with a Glauber model, the rms matter radii of these nuclides has been
extracted \cite{Kanungo16} and is presented in Fig.~\ref{fig:fig1}b. The extracted matter radii are 
compared to the results of relativistic mean field (RMF) \cite{Serot86,Ring96} calculations 
using the NL3 and FSUGOLD interactions. In contrast to the widely used NL3 interaction,
the FSUGOLD corresponds to a softer interaction and consequently results in a slightly larger matter radius. 
It is noteworthy that the NL3 interaction provides a somewhat more accurate description of the extracted radii as compared to the FSUGOLD interaction. 
Although the FSUGOLD interaction provides a slightly larger radius for all isotopes as compared to the NL3 interaction, the dependence on neutron excess is essentially the same.

\begin{figure}
\includegraphics[scale=0.45]{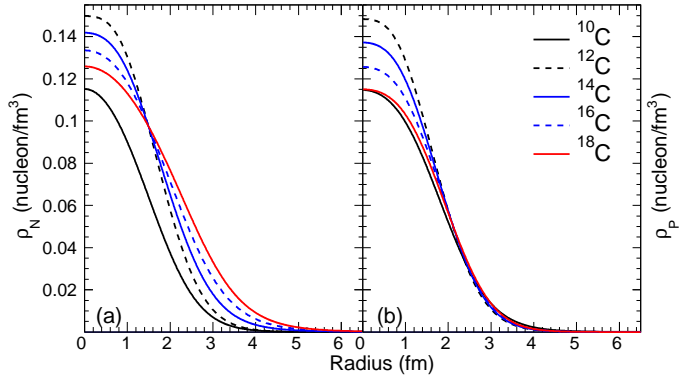}
\caption{\label{fig:fig2} (Color online) Density distributions of neutrons and protons (panels a and b respectively) 
predicted by the RMF model 
for carbon isotopes using the NL3 interaction. 
}
\end{figure}

Presented in Fig.~\ref{fig:fig2} are the neutron and proton density distributions for various carbon isotopes 
predicted by the
RMF model using the NL3 interaction. 
Examination of the neutron density distributions reveals that with increasing neutron number,
the tail of the neutron density distribution extends further as expected.
For the proton density distributions, for N$>$Z, the largest observable change is the decrease in the central density with increasing neutron number. This decrease in the central density is correlated with a slight increase in the tail of the proton distribution due to the attractive nuclear force of the valence neutrons. This influence of the additional neutrons on the  charge radii 
has been experimentally established  for the carbon isotopic chain
through measurement of the charge changing cross-section \cite{Kanungo16}. Distributions calculated with the FSUGOLD interaction exhibit the same trends as those presented.

\begin{figure}
\includegraphics[scale=0.45]{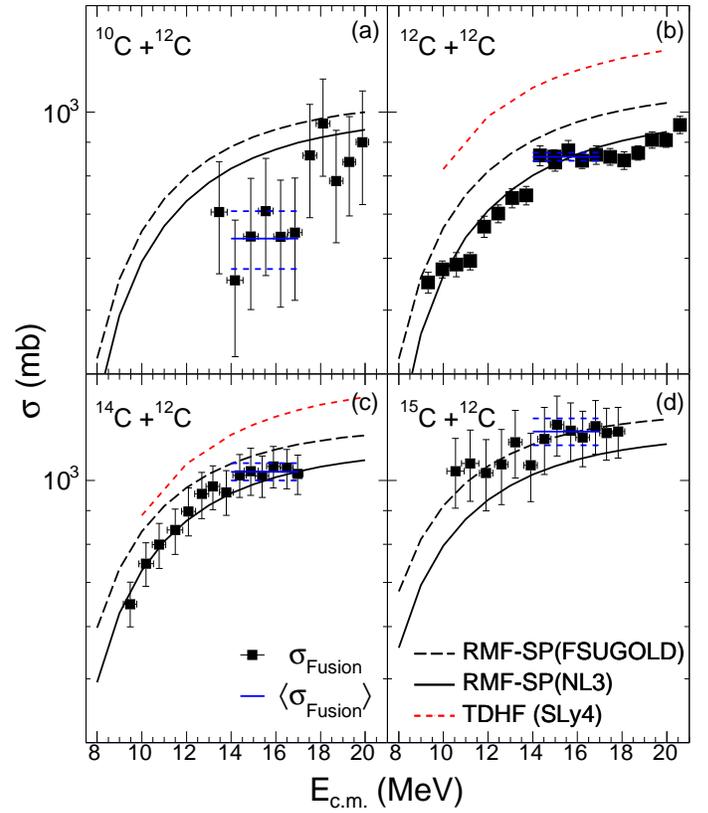}
\caption{\label{fig:fig3} (Color online) Fusion excitation functions for $^{10-15}$C + $^{12}$C. Experimental data are compared with the results of a RMF-SP model using the FSUGOLD and NL3 interactions. Predicted fusion cross-sections with a TDHF model with a SLy4 interaction are indicated.
The fusion excitation function for $^{13}$C (not shown) is comparable to that of $^{14}$C. Data taken from \cite{Calderon15, Carnelli15}.
}
\end{figure}

\begin{figure}
\includegraphics[scale=0.42]{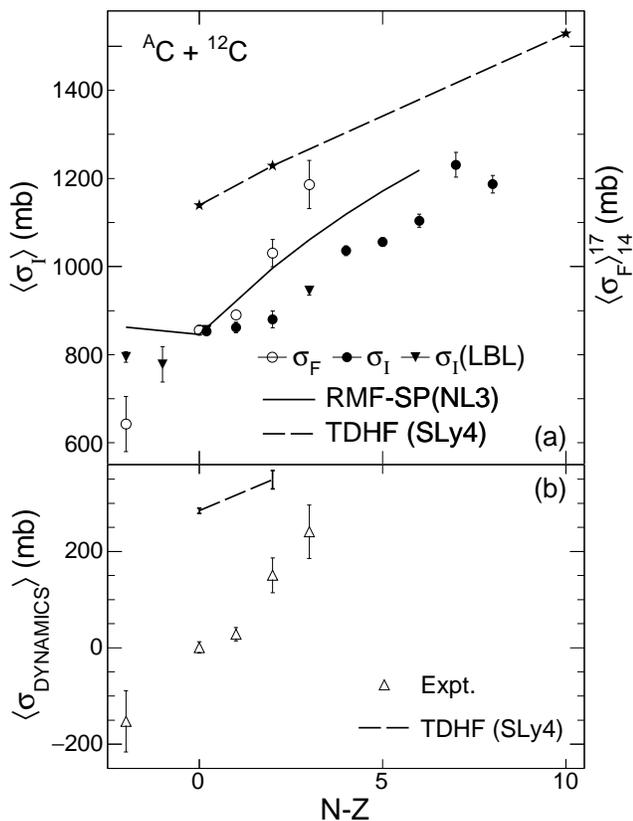}
\caption{\label{fig:fig4} Panel a) Comparison of the dependence of the average above-barrier fusion 
cross-section, $<$$\sigma_{F}$$>$$_{14}^{17}$, and the 
interaction cross-section, $\sigma_{I}$, on neutron excess. Experimental results
are compared with the predictions of the RMF-SP and TDHF models. Panel b) Dependence of the average cross-section due to dynamics on neutron excess.
}
\end{figure}

To examine the response of the neutron and proton density distributions  
to the perturbation involved in a collision we investigate the evolution of the fusion cross-section 
with increasing neutron number. We specifically examine the fusion cross-section for $^A$C+ $^{12}$C 
at near barrier energies where long interaction times allow an adiabatic response of the density distributions. 
Using a novel active target approach the fusion excitation functions for these reactions was
measured by the ANL group \cite{Carnelli14, Carnelli_thesis}. This active target approach is particularly well suited to studying reactions
with low-intensity beams and allowed measurement of the fusion excitation function with
beam intensities as low as 500 ions/s. Depicted in Fig.~\ref{fig:fig3} are the 
cross-section data for $^{10,12,13,15}$C 
that have been taken from \cite{Carnelli15} along with $^{14}$C cross-sections measured
using the same approach but at higher beam intensity beams 
\cite{Calderon15}. While an improved analysis, primarily a reduction of background arising from scattered beam, resulted in publication of revised cross-sections for $^{10,15}$C \cite{Carnelli15, Rehm20}, systematic comparison of the neutron number dependence 
of the fusion cross-section for the isotopic chain has not been published.
The measured fusion excitation
functions for  $^{12,13}$C + $^{12}$C are in good agreement with those 
published in the literature \cite{Kovar79} providing confidence in this approach to  
measure the fusion excitation function. 

Having established that the RMF model provides a reasonably accurate description of the matter radii of the 
carbon isotopes, we utilize the predicted RMF density distributions together with the Sao Paulo (SP) 
model \cite{Gasques04} to 
predict the fusion cross-section. The fusion calculations were performed at energies of E/A = 2-3 MeV and
the results are depicted in Fig.~\ref{fig:fig3} as the solid and long-dashed lines.
While in the case
of $^{10}$C the model overpredicts the experimental results, in the remainder of the cases the agreement is reasonable. Calculations with FSUGOLD consistently predicts larger cross-sections
than those with NL3, consistent with the larger radii for FSUGOLD observed in Fig.~\ref{fig:fig1}b. 
It is noteworthy that this increase in the cross-section for FSUGOLD as compared to NL3 
is typical of the entire above-barrier regime. Although for $^{10}$C and $^{12}$C the RMF-SP(NL3) 
calculations provide a better description of the excitation functions, in the case of $^{15}$C, the 
RMF-SP(NL3) calculations underpredict the measured cross-sections. For 
 $^{15}$C a better description is achieved
using the FSUGOLD interaction. This trend indicates that the RMF-SP for a given interaction, either NL3 
or FSUGOLD, does not exhibit the same dependence on neutron number as the experimental data.
We therefore explore the dependence of the average fusion cross-section on neutron excess.
Depicted in Fig.~\ref{fig:fig3} as the blue bar is the
value of the average experimental cross-section. The bar also indicates the energy interval of
14 MeV$\le$E$_{c.m.}$$\le$17 MeV over which the average was calculated and the resulting quantity is
designated $\langle$$\sigma_{F}$$\rangle$$_{14}^{17}$. 
For $^{12-15}$C the average cross-section is clearly representative of the above-barrier cross-section. 
In the case of $^{10}$C the choice of energy interval could result in an average cross-section that 
is uncharacteristicaly low and therefore unrepresentative of the above-barrier behavior. The upper energy limit of 17 MeV is dictated by limited data for $^{14}$C motivating future measurement.

The average above-barrier fusion cross-sections, $\langle$$\sigma_{F}$$\rangle$$_{14}^{17}$,
are juxtaposed in Fig.~\ref{fig:fig4}a with
the measured interaction cross-sections, $\sigma_{I}$. 
One observes that for $^{12}$C the fusion cross-section and the $\sigma_{I}$ are essentially the same. 
For N$>$Z however, the fusion cross-section depends more strongly on neutron excess 
than $\sigma_{I}$ does. Since the dependence of
$\sigma_{I}$ on increasing neutron number indicated the inherent growth in the size of the 
matter radius with
increasing neutron number, the increased cross-section in the case of fusion 
reflects the impact of dynamics in the fusion process.
Moreover, the larger slope for fusion as compared to $\sigma_{I}$ indicates that this dynamics 
increases with increasing neutron excess.

Evidence for the dependence of the fusion dynamics on neutron number is also evident 
in Fig.~\ref{fig:fig4}a by 
comparing the experimental data with the RMF-SP(NL3) calculations.
For N$\le$Z the value of $<$$\sigma_{F}$$>$$_{14}^{17}$ predicted by the RMF-SP(NL3) model
is essentially constant while for N$>$Z it increases approximately linearly. 
It is clear that the experimental data
exhibit a stronger dependence on neutron excess than the RMF-SP model. 
This comparison reinforces the increasing importance of dynamics with increasing 
neutron excess. 
Calculations with the FSUGOLD interaction (not shown) though yielding a slightly larger cross-section,
exhibit the same increase with neutron excess for N$>$Z. 

A key observation from Fig.~\ref{fig:fig4}a 
is the similarity of the slopes of the RMF-SP(NL3) calculations for N$>$Z and
that for $\sigma_{I}$ with N$\ge$Z. This similarity of the two slopes arises
from the fact that the $\sigma_{I}$ measures the size of the nucleus and the RMF-SP with the frozen density distributions is
intrinsically related to the same quantity. {\em As such, the quantity $\sigma_{I}$ provides a key 
experimental reference from which to examine fusion dynamics.} We therefore present in 
Fig.~\ref{fig:fig4}b the difference between the average above-barrier fusion cross-section
and the interaction cross-section namely, 
$\langle$$\sigma_{DYNAMICS}$$\rangle$
=$\langle$$\sigma_{F}$$\rangle$$_{14}^{17}$ - $\sigma_{I}$. 
This quantity represents the average cross-section due to the fusion dynamics. 
Although the span of neutron excess for the fusion
data presented is presently limited, 
the experimental data manifests a 
linear behavior of the fusion dynamics on neutron excess.

To investigate the dependence of dynamics on neutron number, we performed calculations with
a time-dependent Hartree-Fock (TDHF) model.
On general grounds the TDHF
approach is well-suited to describing the 
large-amplitude collective motion associated with fusion. 
Advances in theoretical and computational techniques 
allow TDHF calculations to be performed on a 3D Cartesian grid thus eliminating 
artificial symmetry restrictions~\cite{Umar06a}. 
Although in the sub-barrier regime in order to accurately describe the 
fusion cross-sections it is necessary to perform density 
constrained TDHF (DC-TDHF) calculations~\cite{Umar12, deSouza13, Steinbach14a}
to obtain the heavy-ion potentials \cite{Umar06d},
at the above-barrier energies
considered in this work direct TDHF calculations can be performed by initiating 
collisions
for increasing impact parameters until the maximum impact parameter for fusion is reached.
In practice this was done with an impact parameter precision of 0.01~fm. Calculations were
performed using the TDHF model, Sky3D \cite{Sky3D} with a SLy4 interaction \cite{Chabanat97, Douchin01}. 
Fusion
calculations were performed for collision of $^{12,14,22}$C projectiles with a $^{12}$C target. 
Due to the sensitivity of the fusion cross-section to pairing \cite{Steinbach14a}, 
exacerbated for odd-A nuclei, and the deformed ground states of some even-A isotopes, calculations were restricted to these cases.

Presented in Fig.~\ref{fig:fig1}b as the open squares are the RMS matter radii predicted by the 
TDHF model for $^{12}$C and $^{14}$C.
The radii predicted by the TDHF model are in reasonable agreement with the RMF calculations and only 
slightly 
larger than those extracted experimentally. The fusion excitation functions predicted by TDHF for 
$^{12}$C+$^{12}$C and $^{14}$C+$^{12}$C are shown in 
Fig.~\ref{fig:fig3} as the short dashed line. 
The TDHF model systematically overpredicts the measured fusion excitation functions. 
This overprediction has been previously reported for similar systems \cite{Steinbach14a, Carnelli14}. More interesting is the 
dependence of the average fusion cross-section predicted by TDHF on neutron excess evident in 
Fig.~\ref{fig:fig4}. A linear behavior of the TDHF predicted cross-section on neutron excess is 
observed indicating that neither any additional enhancement or supression of dynamics is predicted 
despite the extreme neutron-richness of $^{22}$C.
This linear behavior manifests the same slope as the dependence of $\sigma_{I}$ on neutron excess 
defined by $^{12,16,17,18,20}$C indicating that for the TDHF model dynamics provides essentially a 
constant increase to the cross-section above the geometric size. The magnitude of this increase due to the dynamics in the TDHF model is $\approx$280 mb. The larger value of $\sigma_{I}$ for the weakly-bound 
nucleus $^{19}$C has been associated with the halo nature of its unpaired neutron \cite{Kanungo16}.
Due to the absence of data for $\sigma_{I}$ in the case of $^{22}$C, $\langle$$\sigma_{DYNAMICS}$$\rangle$
can only be calculated in the TDHF model for $^{12}$C and $^{14}$C. In Fig.~\ref{fig:fig1}b 
the dependence of $\langle$$\sigma_{DYNAMICS}$$\rangle$ on neutron excess for the TDHF model arises purely 
from the near constancy of $\sigma_{I}$, a consequence of shell structure.
The strong dependence of 
$\langle$$\sigma_{DYNAMICS}$$\rangle$ 
on neutron excess clearly exhibited in Fig.~\ref{fig:fig4}b by the experimental data 
is not described by the TDHF model. This result suggests that an important aspect of the 
dynamics is not incorporated in the model.

Examination of the fusion cross-section at above-barrier energies for an isotopic chain is a 
powerful tool for investigating nuclear dynamics.
Comparison of average fusion cross-sections just above the barrier with
the interaction cross-section, $\sigma_{I}$, 
at high energies where the sudden approximation 
is valid allows extraction of not just
the fusion dynamics but the dependence of the dynamics on neutron excess. For the light nuclei considered in this work, a widely-accepted
dynamical model of fusion, namely a time-dependent Hartee-Fock model, fails to describe the dependence of the dynamics on neutron excess.
By investigating this dynamics for the most neutron-rich
nuclei, valuable insight into the dynamics of extremely asymmetric nuclear matter can be gained.

\begin{acknowledgments}
We gratefully acknowledge helpful discussions on both the general topic of fusion as well as 
the TDHF model with S. Umar (Vanderbilt University).
This work was supported by the U.S. Department of Energy under Grant No. 
DE-FG02-88ER-40404 (Indiana University).
CJH is supported in part by U.S. DOE grants DE-FG02-87ER40365 and DE-SC0018083.
ZL gratefully acknowledges support from National Science Foundation under PHY-1613708 and DOE grant
DE-SC0019470 (Arizona State University).
\end{acknowledgments}



%

\end{document}